\documentclass[11pt]{article}
\usepackage{amsmath,amssymb,color}

\def\ben{\begin{equation}}
\def\een{\end{equation}}

\def\nn{\nonumber} \def\bd{\begin{document}} \def\ed{\end{document}}
\def\ds{\documentstyle} \let\fr=\frac \let\bl=\bigl \let\br=\bigr
\let\Br=\Bigr \let\Bl=\Bigl
\let\bm=\bibitem
\let\na=\nabla
\let\pa=\partial \let\ov=\overline
\newcommand{\be}{\begin{equation}}
\newcommand{\ee}{\end{equation}}
\def\ba{\begin{array}}
\def\ea{\end{array}}
\def\ft#1#2{{\textstyle{\frac{\scriptstyle #1}{\scriptstyle #2} } }}
\def\fft#1#2{{\frac{#1}{#2}}}
\def\del{\partial}
\def\vp{\varphi}
\def\sst#1{{\scriptscriptstyle #1}}
\def\oneone{\rlap 1\mkern4mu{\rm l}}
\def\td{\tilde}
\def\wtd{\widetilde}
\def\ie{{\it i.e.\ }}
\def\dalemb#1#2{{\vbox{\hrule height .#2pt
        \hbox{\vrule width.#2pt height#1pt \kern#1pt
                \vrule width.#2pt}
        \hrule height.#2pt}}}
\def\square{\mathord{\dalemb{6.8}{7}\hbox{\hskip1pt}}}
\def\i{{\rm i}}
\newcommand{\ho}[1]{$\, ^{#1}$}
\newcommand{\hoch}[1]{$\, ^{#1}$}
\newcommand{\bea}{\setlength\arraycolsep{2pt} \begin{eqnarray}}
\newcommand{\eea}{\end{eqnarray}}
\newcommand{\ra}{\rightarrow}
\newcommand{\lra}{\longrightarrow}
\newcommand{\Lra}{\Leftrightarrow}
\newcommand{\bp}{\tilde \beta^\prime}
\newcommand{\tr}{{\rm tr} }
\newcommand{\Tr}{{\rm Tr} }
\def\0{{\sst{(0)}}}
\def\1{{\sst{(1)}}}
\def\2{{\sst{(2)}}}
\def\3{{\sst{(3)}}}
\def\4{{\sst{(4)}}}
\def\5{{\sst{(5)}}}
\def\6{{\sst{(6)}}}
\def\7{{\sst{(7)}}}
\def\8{{\sst{(8)}}}
\def\m{{\sst{(m)}}}
\def\n{{\sst{(n)}}}
\def\cA{{{\cal A}}}
\def\cB{{{\cal B}}}
\def\cF{{{\cal F}}}
\def\cG{{{\cal G}}}
\def\cH{{{\cal H}}}
\def\tV{\widetilde V}
\def\tW{\widetilde W}
\def\tH{\widetilde H}
\def\tE{\widetilde E}
\def\tF{\widetilde F}
\def\tA{\widetilde A}
\def\im{{{\rm i}}}
\def\tY{{{\wtd Y}}}
\def\ep{{\epsilon}}
\def\vep{{\varepsilon}}
\def\bD{{{\bar D}}}
\def\R{{{\mathbb R}}}
\def\C{{{\mathbb C}}}
\def\H{{{\mathbb H}}}
\def\CP{{{\mathbb C}{\mathbb P}}}
\def\RP{{{\mathbb R}{\mathbb P}}}
\def\Z{{{\mathbb Z}}}
\def\bA{{{\mathbb A}}}
\def\bB{{{\mathbb B}}}
\def\bC{{{\mathbb C}}}
\def\bD{{{\mathbb D}}}
\def\bE{{{\mathbb E}}}
\def\bZ{{{\mathbb Z}}}
\def\Re{{{\frak{Re}}}}
\def\Im{{{\frak{Im}}}}
\def\cosec{{\,\hbox{cosec}\,}}
\def\Gm{{\Gamma_{\!\! -}}}
\def\Gp{{\Gamma_{\!\! +}}}
\def\stan{{standard }}
\def\nonstan{{supernumerary }}
\def\p{{\partial}}
\def\kdel#1{{\fft{\del}{\del#1}}}

\def\bog{{Bogomolny }}
\def\om{{\omega}}

\newcommand{\nnr}{\nonumber \\}
\newcommand{\pd}{\partial}
\newcommand{\ud}{\textrm{d}}
\newcommand{\dTH}{T^{\prime \, 0}_\textrm{H}}
\newcommand{\dOi}{\Omega^{\prime \, 0}_i}
\newcommand{\bx}{{\bf x}}

 \newcommand{\bpsi}{\bar\psi}
 \newcommand{\cL}{{\cal L}}
 \newcommand{\cR}{{\cal R}}
 \newcommand{\cT}{{\cal T}}
 \def\bra#1{\left\langle#1\right|}
 \def\ket#1{\left|#1\right\rangle}
 \def\brk#1{\left\langle#1\right\rangle}

\newcommand{\tamphys}{\it George and Cynthia Woods Mitchell  Institute
for Fundamental Physics and Astronomy,\\
Texas A\&M University, College Station, TX 77843-4242, USA}

\newcommand{\auth}{H. L\"u\hoch{\dagger}, Yi Pang\hoch{\ddagger} and Zhao-Long Wang\hoch{\star}}

\begin{document}

\begin{flushright}
\hfill{
USTC-ICTS-09-19\ }
\end{flushright}

\vspace{25pt}
\begin{center}
{\large {\bf Constructing Calabi-Yau Metrics From Hyperk\"ahler
Spaces}}

\vspace{15pt}

\auth

\vspace{10pt}

\hoch{\dagger}{\it China Economics and Management Academy\\
Central University of Finance and Economics, Beijing, 100081, China}

\vspace{10pt}

\hoch{\ddagger}{\it Key Laboratory of Frontiers in Theoretical
Physics\\
Institute of Theoretical Physics, Chinese Academy of Sciences,
Beijing 100190}

\vspace{10pt}

\hoch{\star}{\it Interdisciplinary Center for Theoretical Study,\\
University of Science and Technology of China, Hefei, Anhui
230026, China}

\vspace{40pt}

\underline{ABSTRACT}
\end{center}

Recently, a metric construction for the Calabi-Yau 3-folds from a
four-dimensional hyperk\"ahler space by adding a complex line bundle
was proposed. We extend the construction by adding a $U(1)$ factor
to the holomorphic $(3,0)$-form, and obtain the explicit formalism
for a generic hyperk\"ahler base.  We find that a discrete choice
arises: the $U(1)$ factor can either depend solely on the fibre
coordinates or vanish. In each case, the metric is determined by one
differential equation for the modified K\"ahler potential. As
explicit examples, we obtain the generalized resolutions (up to
orbifold singularity) of the cone of the Einstein-Sasaki spaces
$Y^{p,q}$. We also obtain a large class of new singular CY3 metrics
with $SU(2)\times U(1)$ or $SU(2)\times U(1)^2$ isometries.

\vspace{15pt}

\thispagestyle{empty}





\newpage

\section{Introduction}

      Six-dimensional Calabi-Yau manifolds (CY3)
\cite{calabi1,calabi2,styau} play an important role in string
theory, since they provide natural internal compactifying spaces,
giving rise to four-dimensional theories that preserve one quarter
of the ten-dimensional supersymmetry.  String compactification on
the CY3 has been studied mainly based on their topological
properties.  This is because although it was demonstrated that
Ricci-flat and complete metrics exist on the compact CY3 manifolds
\cite{styau}, one does not expect to see an explicit one, aside from
the flat metric. This can be seen from the Killing vector analysis.
A Killing vector $K_i$ satisfies the equation
\be
           -\Box K_i - R_{ij} K^j = 0\,.
\ee
Multiply by $K^i$ and integrate over the manifold. For a compact
manifold, integration by parts on the first term gives no boundary
contribution, and hence one concludes
\be \int_{\cal M} ( |\nabla_i K_j|^2 - R_{ij} K^i K^j) =0\,. \ee
For the Ricci-flat CY3 metrics, it must be that $\nabla_i K_j = 0$
pointwise everywhere in the manifold.  Leaving aside the trivial
possibility that there are flat $S^1$ factors, such a
covariantly-constant vector will not exist. Therefore there can be
no Killing vectors in a non-flat Ricci-flat compact manifold.
Without Killing vectors, there are no continuous symmetries.  And
without the simplifications that result from supposing that a metric
has continuous symmetries, it is essentially hopeless to solve the
Einstein equations.

     Thus explicit metrics on the CY3 with Killing
vectors are necessarily non-compact and/or singular.  In fact, in
string theory compactification, the CY3 spaces can develop
singularities at limiting values of their modulus parameters, where
additional massless four-dimensional states will emerge. The
simplest example of such a singular metric is the conifold, which is
the Ricci-flat metric on the cone over the homogeneous
Einstein-Sasaki space $T^{1,1} = (S^3\times S^3)/U(1)_{\rm diag}$.
The metric is singular at the vertex of the cone, and as the moduli
are moved slightly away from the singular limit, the metric near to
the previous conifold point is then smoothed out.  It was shown
\cite{conifolds} that there are two possible ways of smoothing out
the conifold: in one of which the vertex is blown up to an $S^2$; in
the other, it is blown up to an $S^3$.

    Not so long ago, it was demonstrated by explicit construction that
there exist an infinite number of Einstein-Sasaki metrics with the
toric $U(1)^3$ isometry, called $Y^{p,q}$ \cite{ypq} and $L^{p,q,r}$
\cite{lpqr1,lpqr2}.  These provide an infinite number of generalized
(toric) conifolds, some of which can be smoothed out completely
\cite{ooya,lpres,cclpv} or up to an orbifold singularity
\cite{lpres,cclpv}.  The construction can be generalized to
arbitrary $2n$ dimensions, since it turns out that the local metrics
can be obtained directly from the BPS limits
\cite{hsy,lpqr1,clpkerr} of the Euclidean version of the general
Kerr-AdS black hole \cite{hht,glppbh1,glppbh2} and Kerr-AdS-NUT
solutions \cite{clpkerr} in arbitrary dimensions: the odd
dimensional ones give rise to Einstein-Sasaki spaces whilst the even
dimensional ones give rise to Calabi-Yau metrics.

      The same Killing vector analysis applies to the higher
dimensional manifolds with reduced holonomy and one does not expect
to see an explicit metric on compact spaces. Examples of complete
metrics on non-compact manifolds in higher dimensions with $G_2$ and
$spin(7)$ holonomy were constructed in the later 1980's
\cite{bb,pp,brsa,gpp}. Inspired by the AdS/CFT correspondence and
applications in M-theory compactification, large classes of explicit
metrics on $G_2$ and $spin(7)$ holonomy spaces have been constructed
and their applications in string and M-theory have been discussed
\cite{com1}-\cite{com11}.

    Although a large number of Calabi-Yau metrics have been
constructed, an organizing principle is still lacking, since many of
these metrics are discovered serendipitously, or constructed
indirectly through the BPS limit of the known Kerr-AdS-NUT
solutions. Recently, a new technique was developed in \cite{sant}
for constructing CY3 metrics, generalizing the construction of the
D6-brane wrapping on a two-cycle of a four-dimensional hyperk\"ahler
space \cite{fay}.  The essence of the construction is to build a CY3
metric from a hyperk\"ahler one by adding a complex line bundle.
This follows the same line of constructing $(2n+2)$-dimensional
Einstein-K\"ahler spaces from $(2n)$-dimensional ones \cite{pp}.
Differently, in the new construction, the K\"ahler potential in four
dimensions is allowed to be modified by an arbitrary function $G$.
However, the proof of the existence of such CY3 metrics were
presented for the $\R^4$ base only.  What is curious is that the
metrics with the asymptotic structure of cones over Einstein-Sasaki
spaces, such as the $\R^6$ or conifolds, are absent from the
construction presented in \cite{sant}.

  In section 2, we extend this construction by considering a generic
hyperk\"ahler base.  Furthermore, we find that the ansatz for the
holomorphic $(3,0)$-form presented in \cite{sant} can be
supplemented with a $U(1)$ factor. This allows us to construct a
much wider class of solutions including ones that have asymptotic
cones over Einstein-Sasaki spaces.  There are two discrete
possibilities for the $U(1)$ factor. One is that it is dependent
solely on the fibre coordinates.  In this case the equations for the
CY3 are reduced to one differential equation for the modified
K\"ahler potential $G$. The other possibility is that the $U(1)$
factor vanishes, for which the metric is determined by a
differential equation which is the singular limiting case of the
previous one. In section 3, we consider the simplest $\R^4$ base and
obtain the CY3 metrics that describe resolutions of the cone over
the $Y^{p,q}$ spaces, when the $U(1)$ factor in the $(3,0)$-form is
non-vanishing. For the case with vanishing $U(1)$ factor, we
obtained a large class of new singular metrics. In section 4, we
consider triaxial basis for the hyperk\"ahler spaces that preserve
the $SU(2)$ isometry. Again solutions with or without the $U(1)$
factors were obtained. We conclude our paper in section 5.
In appendices A and B, we present some complicate formulae and detailed
derivations.

\section{The construction}

\subsection{The ansatz}

     In this section, we review and then extend significantly
the metric construction for the CY3 from $D=4$ hyperk\"ahler spaces,
proposed in \cite{sant}. Let us consider a generic hyperk\"ahler
space in four dimensions with the complex coordinates $z^i,\bar
z^i~(i=1,2)$ and the K\"ahler potential $K_0(z^i,\bar z^i)$, The
metric is given by
\bea ds^2=2\tilde g_{i\bar j} dz^i d{\bar z}^j\,,\qquad\tilde
g_{i\bar j}=\fft12\partial_i{\partial}_{\bar j}K_0=\tilde g_{\bar j i}\,.
\eea 
Since it is Ricci flat, the Ricci form ${\mathcal R}^{(1,1)}$
vanishes, {\it i.e.}
\begin{equation}\label{Ricci}
{\mathcal R}^{(1,1)}=i\tilde\partial{\bar{\tilde\partial}}
\log\sqrt{V}=0\,.
\end{equation}
Here $V\equiv{\rm det}(\tilde g_{i\bar j})^2$ is the volume factor
and $\tilde \partial$ and $\bar{\tilde\partial}$ are the Dolbeault
1-form differential operators defined by
\be \tilde \partial \equiv dz^i\, \partial_{z^i}\,,\qquad
\bar{\tilde\partial} \equiv d{\bar z^i}\,\partial_{\bar z^j}\,. \ee
The equation (\ref{Ricci}) implies that $\log {V}$ is the real part
of a holomorphic function, or equivalently, $V$ can be the norm of a
holomorphic function. The choice for the complex coordinates is not
unique since we can always make a holomorphic coordinate
transformation $z^i\rightarrow z'^i=f^i(z^j)$. Under such a
transformation, the volume factor transforms as
\begin{equation}
V\rightarrow |T|^{-4}V\,,
\end{equation}
where
\begin{equation} T={\rm det}\left[\frac{\partial(f^1,f^2)}
{\partial (z^1,z^2)}\right]\,.
\end{equation}
It is easy to see that $T$ can be any holomorphic function. Thus we
can always set $V=1$ by choosing appropriate complex coordinates. We
shall do this for later convenience.

    We now consider a complex vielbein basis of the hyperk\"ahler
space, which is given by $\tilde\epsilon^a, \bar{\tilde\epsilon}^a~
(a=1,2)$, then the corresponding metric, K\"ahler form and
holomorphic $(2,0)$-form are given by
\bea ds^2&=&\tilde \epsilon^1\bar {\tilde\epsilon}^1+\tilde
\epsilon^2\bar {\tilde\epsilon}^2\,,\cr \tilde J^{(1,1)}&=&\frac{\rm
i}{2}(\tilde\epsilon^1\wedge\bar{\tilde\epsilon}^1+
\tilde\epsilon^2\wedge\bar{\tilde\epsilon}^2)\,,\cr
\tilde\Omega^{(2,0)}&=&\tilde\epsilon^1\wedge\tilde\epsilon^2\,.
\eea 
According to the properties of hyperk\"ahler spaces, both $\tilde J$
and $\tilde\Omega$ are closed, {\it i.e.}\ $d_4\tilde
J=d_4\tilde\Omega=0$. One can now use this structure to construct a
CY3.  The metric ansatz is given by \cite{sant}
\bea\label{h}
ds^2&=&ds^2_4+h^2\;dy^2+h^{-2}(d\alpha+A)^2\cr&=&(\delta_{a\bar
b}+G_{a\bar b})\tilde \epsilon^a\bar
{\tilde\epsilon}^b+h^2\;dy^2+h^{-2}(d\alpha+A)^2\;. \eea
The metric components are the functions of the $y,z^i,\bar z^i$
coordinates, but are independent of the coordinate $\alpha$, which
is a manifest Killing direction. Thus the metric ansatz assumes at
least one Killing direction.  The functions appearing in the
$ds^2_4$ part are defined by
\bea \tilde\partial\bar{\tilde\partial}G&=&dz^i\wedge d\bar
z^j\;\partial_{i}\partial_{\bar
j}G=\tilde\epsilon^a\wedge\bar{\tilde\epsilon}^b\;
{\tilde\epsilon}_a^i\bar{\tilde\epsilon}_b^j
\partial_{i}\partial_{\bar j}G =G_{a \bar b}\,{\tilde\epsilon}^a
\wedge\bar{\tilde\epsilon}^b\;,\label{Gabdef} \eea
where ${\tilde\epsilon}_a$ is the inverse complex vielbein. Note
that if we replace $G$ by $K_0$ in (\ref{Gabdef}), we have
$\delta_{a\bar b}$ instead of $G_{a\bar b}$.  Thus the $ds_4^2$ in
(\ref{h}) is obtained by deforming the original K\"ahler potential
$K_0(z^i,\bar z^i)$ to $K_0(z^i,\bar z^i)+G(y,z^i,\bar z^i)$.

   The $ds_4^2$ can be diagonalized by a local $SU(2)$ transformation
$U_a^{~b}(z^i,\bar z^i)$, namely 
\bea U\,  \begin{pmatrix} 1+G_{1\bar1} & G_{1\bar2}
  \cr G_{2\bar1} & 1+G_{2\bar2}
 \end{pmatrix}\, U^{\dag}=\begin{pmatrix} \lambda_1(z^i,\bar z^i) & 0
  \cr 0 & \lambda_2(z^i,\bar z^i)
 \end{pmatrix}\,.
\eea 
We further suppose that the complex structure of the hyperk\"ahler
base is part of complex structure of the CY3. This implies that the
complex vielbein of the CY3 is given by
\bea \epsilon^1=e^{\i\ft\kappa2}\sqrt{\lambda_1} \tilde\epsilon^a
(U^{\dag})_a^{~1}\,,~~~\epsilon^2=e^{\i\ft\kappa2}\sqrt{\lambda_2}
\tilde\epsilon^a (U^{\dag})_a^{~2}\,,~~~
\epsilon^3=h\,dy+\i\,h^{-1}(d\alpha+A)\,. \eea
where $\kappa=\kappa(\alpha,y,z^i,\bar z^i)$ is a real function.
Correspondingly, the K\"ahler form and the $(3,0)$-form for the CY3
are given by 
\bea J^{(1,1)}&=&\frac{\rm
i}{2}(\epsilon^1\wedge\bar{\epsilon}^1+
\epsilon^2\wedge\bar{\epsilon}^2+
\epsilon^3\wedge\bar{\epsilon}^3)
=\fft{\rm i}{2}(\delta_{a\bar b}+G_{a\bar b})
\tilde \epsilon^a\wedge\bar {\tilde\epsilon}^b
+dy\wedge(d\alpha+A)\;,\\
\Omega^{(3,0)}&=&\epsilon^1\wedge\epsilon^2\wedge\epsilon^3
=f\,e^{\i\,\kappa}\,\tilde\epsilon^1\wedge\tilde\epsilon^2
\wedge\left(h\,dy+{\rm i}\,h^{-1}\,(d\alpha+A)\right)\,, \eea
where 
\be\label{f} f=\sqrt{\lambda_1\,\lambda_2}= \sqrt{\det(\delta_{a\bar
b} + G_{a\bar b})} =\sqrt{1+G_{1\bar 1}+G_{2\bar 2}+G_{1\bar
1}G_{2\bar 2}-G_{1\bar 2}G_{2\bar 1}}\,. \ee 
Note that comparing with the ansatz in \cite{sant}, we have
introduced a $U(1)$ factor $e^{{\rm i}\, \kappa}$ for the
holomorphic $(3,0)$-form.  This factor turns out to be crucial for
constructing metrics with asymptotic cones over Einstein-Sasaki
spaces. Although the metric is independent of the coordinate
$\alpha$, this $U(1)$ factor can be.

      The requirement that the metric (\ref{h}) be Calabi-Yau
becomes the requirement that the above K\"ahler form and
$(3,0)$-form are both closed.  Note that
\bea dJ&=&\fft{\rm i}{2}\partial_y G_{a\bar b}\;dy\wedge\tilde
\epsilon^a\wedge\bar {\tilde\epsilon}^b-dy\wedge d_4A =\fft{\rm
i}{2}dy\wedge\tilde\partial\bar{\tilde\partial}\partial_y G
-dy\wedge d_4A \cr&=&-\fft{\rm
i}{4}dy\wedge(\tilde\partial+\bar{\tilde\partial})
(\partial-\bar{\tilde\partial})\partial_y G-dy\wedge d_4A =-\fft{\rm
i}{4}dy\wedge d_4(\partial-\bar{\tilde\partial})\partial_y
G-dy\wedge d_4A\;, \eea
then $dJ=0$ implies that
\be A=-\frac{\rm
i}{4}(\tilde\partial-\bar{\tilde\partial})\partial_y G\,
+\lambda(y,z_i,\bar z_i)\,dy\,, \ee 
up to some pure gauge terms. Note that $d_4$ denotes an exterior
derivative with respect to $z^i$ and $\bar z^i$ only.  The exterior
derivative for the $(3,0)$-form is given by 
\bea d\Omega
&=&\bar{\tilde\partial}(f\,e^{\i\,\kappa}\,h)
\wedge\tilde\epsilon^1\wedge\tilde\epsilon^2\wedge dy +\,{\rm
i}\,f\,e^{\i\,\kappa}\,h\,\partial_\alpha\kappa\,
d\alpha\wedge\tilde\epsilon^1\wedge\tilde\epsilon^2\wedge dy
\cr&&+\,{\rm i}\,\bar{\tilde\partial}(f\,e^{\i\,\kappa}\,h^{-1})
\wedge\tilde\epsilon^1\wedge\tilde\epsilon^2\wedge \left(d\alpha+
\ft{\rm i}{4}\bar{\tilde\partial}(\partial_y
G)+\lambda\,dy\right)\cr &&+\,{\rm i}\,\partial_y
(f\,e^{\i\,\kappa}\,h^{-1})\,dy\wedge\tilde\epsilon^1
\wedge\tilde\epsilon^2\wedge \left(d\alpha+ \ft{\rm
i}{4}\bar{\tilde\partial}(\partial_y G)\right)\cr && -{\rm
i}\,f\,e^{\i\,\kappa}\,h^{-1}\,\tilde\epsilon^1\wedge
\tilde\epsilon^2\wedge dy\wedge\bar{\tilde\partial}\lambda
-\ft14\,f\,e^{\i\,\kappa}\,h^{-1}\,dy\wedge\tilde\epsilon^1
\wedge\tilde\epsilon^2\wedge \bar{\tilde\partial}\partial^2_y G
\cr&&-\,f\,e^{\i\,\kappa}\,h^{-1}\,\partial_\alpha\kappa\,
d\alpha\wedge\tilde\epsilon^1\wedge\tilde\epsilon^2\wedge
\left(\ft{\rm i}{4}\bar{\tilde\partial}(\partial_y
G)+\lambda\,dy\right)\,.\eea
The vanishing of the terms containing $d\alpha\wedge dy$ implies
that 
\bea \label{eq1}\partial_y
(f\,h^{-1})-f\,h\,\partial_\alpha\kappa&=&0\,,\\ \label{eq2}
\partial_y \kappa-\lambda\,\partial_\alpha\kappa&=&0\,.\eea 
Since by construction only $\kappa$ can depend on $\alpha$, it
follows from (\ref{eq1}) that we have
\be \kappa=\alpha\,\kappa_1(y,z^i,\bar z^i)+\kappa_0(y,z^i,\bar
z^i)\,. \ee
Substituting it back to (\ref{eq2}), we find 
 \bea\label{ay} &&\kappa_1\lambda=\partial_y\kappa_{0}\,,~~~~
\kappa=\alpha\,\kappa_1(z^i,\bar z^i)+\kappa_0(y,z^i,\bar
z^i)\,,\\
\label{fh} &&\partial_y g=\kappa_1\,g^{-1}\,f^2\,,\eea
where
\be g \equiv f\,h^{-1}.\ee 
The vanishing of the other terms containing $d\alpha$ implies that
\bea\label{fh-1} &&4\bar{\tilde\epsilon}_a^{\mu}\partial_{\mu}
(g\,e^{\i\,\kappa})
+g\,e^{\i\,\kappa}\,\kappa_1\,\bar{\tilde\epsilon}_a^{\mu}
\partial_{\mu}\partial_y G=0\,, \eea
from which, we find
\bea\label{k1} \bar{\tilde\epsilon}_a^{\mu}\partial_{\mu}
\kappa_1=0\,,\\\label{fg-}
\bar{\tilde\epsilon}_a^{\mu}\partial_{\mu}\left(
4\,\log g+\i\,4\,\kappa_0
+\kappa_1\,
\partial_y G\right)=0\,.\eea 
Since $\kappa_1$ is real, (\ref{k1}) implies that $\kappa_1$ is a
constant.  It can be set to either 0, or 1, by rescaling the
$\alpha$ coordinate. The vanishing of the rest terms implies that 
\bea &&4\bar{\tilde\epsilon}_a^{\mu}\partial_{\mu}
\left(g^{-1}\,f^2\,e^{\i\,\kappa}+\i\,g\,e^{\i\,\kappa}\,\lambda\right)
+\partial_y\left(g\,e^{\i\,\kappa}\,
\bar{\tilde\epsilon}_a^{\mu}\partial_{\mu}
\partial_y G\right)=0\,,
\cr&\Rightarrow&\,
\bar{\tilde\epsilon}_a^{\mu}\partial_{\mu}\left(4\,g^{-2}
\,f^2+\i\,4\,\lambda +\partial^2_yG\right)=0\,,\label{fg} \eea 
where we have used the equations (\ref{ay}), (\ref{fh}),
(\ref{fh-1}) and (\ref{fg-}).

\subsection{Case I: $\kappa_1=1$}

For $\kappa_1=1$, the equation (\ref{fg}) can be deduced from
(\ref{fh}) and (\ref{fg-}). Then the CY3 is determined by the
following equations
\bea &&\label{general1}g\,\partial_y g=1+G_{1\bar
1}+G_{2\bar 2}+G_{1\bar
1}G_{2\bar 2}-G_{1\bar 2}G_{2\bar 1}\,,\\
&&\label{general2}\bar{\tilde\epsilon}_a^{\mu}\partial_{\mu}\left(
\partial_y G+4\,\log g+\i\,4\,\kappa_0\right)=0\,,
\eea
and the other quantities are then given by
\be
\lambda=\partial_y\kappa_{0}\,,~~~\kappa=\alpha+\kappa_0(y,z^i,\bar
z^i)\,. \ee 
Note that a partial derivative of (\ref{general2}) with respect to
$y$ gives rise to (\ref{fg}).  The equation (\ref{general2}) implies
that we
have $\partial_y G+4\log g+\i\,4\,\kappa_0 =H(y,z^i)$, where 
\be H\equiv U + \i\, V \label{Hfun}\ee
is any holomorphic function on the hyperk\"ahler base. The real part
$U=\partial_y G + 4 \log g$ satisfies 
\be\label{Geq1} \tilde\epsilon_a^{\mu}\bar{\tilde\epsilon}_b^{\nu}
(\partial_{\mu}\partial_{\nu}-\Gamma_{\mu\nu}^\lambda
\partial_{\lambda})
U=0\,,\ee
{The imaginary part is then given by
\be
4\,\kappa_0=V=\int({\tilde\partial}+\bar{\tilde\partial})V
=-\i\,\int({\tilde\partial}-\bar{\tilde\partial})
U\,. \ee
However, note that $\partial_yG\rightarrow \partial_yG-U$ and
$\alpha\rightarrow \alpha+\kappa_0$ are gauge transformations in our
set up. Therefore, we can always set $H=0$. Then we have
$\kappa_0=\lambda=0$, and that 
\be g=\exp\left(-\fft{1}4\partial_y G\right).\ee
It follows from (\ref{general1}) that the system will be determined
solely by the following basic equation for $G$ 
\be
\partial_y \left[\exp\left(-\fft{1}2\partial_y G\right)\right]
=2(1+G_{1\bar 1}+G_{2\bar 2}+G_{1\bar 1}G_{2\bar 2}-G_{1\bar
2}G_{2\bar 1})\,.\label{basicG1} \ee 
The $U(1)$ factor depends on the fibre coordinate $\alpha$ only.

\subsection{Case II: $\kappa_1=0$}

When $\kappa_1=0$, the CY3 is determined by the
following equations
\bea \label{general3}&&\bar{\tilde\epsilon}_a^{\mu}\partial_{\mu}
\left( \log g+\i\,\kappa_0\right)=0\,,\\
&&\label{general4}\bar{\tilde\epsilon}_a^{\mu}\partial_{\mu}
\left[\partial^2_yG+4\,g^{-2}\,(1+G_{1\bar 1}+G_{2\bar 2}+G_{1\bar
1}G_{2\bar 2}-G_{1\bar 2}G_{2\bar 1})+\i\,4\,\lambda \right]=0\,,
\eea
where $\kappa_0=\kappa_0(z^i,\bar z^i)$ and $g=g(z^i,\bar z^i)$.
It follows that both
\be g\,e^{\i\,\kappa_0}=H_1(z^i)\ee 
and
\be\partial^2_yG+4\,g^{-2}\,(1+G_{1\bar 1}+G_{2\bar 2}+G_{1\bar
1}G_{2\bar 2}-G_{1\bar 2}G_{2\bar 1})+ \i\,4\,\lambda=
H_2(y,z^i)=U+\i\,V \ee
are holomorphic functions on the hyperk\"ahler base. Again, the
gauge transformations $\partial^2_yG\rightarrow \partial^2_yG-U$ and
$\alpha\rightarrow \alpha+\Lambda$ imply that we can always set
$H_2=0$. Then we have $\lambda=0$ and
\be\partial^2_yG+4\,g^{-2}\,(1+G_{1\bar 1}+G_{2\bar 2}+G_{1\bar
1}G_{2\bar 2}-G_{1\bar 2}G_{2\bar 1})=0\,.\label{basicG2.1}
 \ee
Furthermore, let us consider a holomorphic coordinate transformation
$z^i\rightarrow \omega^i(z^j)$. It induces the following
transformation on the four dimensional K\"ahler potential
\bea G\rightarrow\tilde G(y,z^i,\bar z^i)=K_0(\omega^i,\bar
\omega^i)-K_0(z^i,\bar z^i)+G(y,\omega^i,\bar\omega^i)\,, \eea
where $K_0$ is the K\"ahler potential for the hyperk\"ahler base. If
the holomorphic functions $\omega^i(z^j )$ satisfy 
\be\label{holotrans} {\rm det}\left[\frac{\partial (z^1,z^2)}{\partial(
\omega^1,\omega^2)}\right]=H_1(\omega^i)\,,\ee 
it follows from (\ref{f}) that the equation (\ref{basicG2.1})
becomes 
\be
\partial^2_y \tilde G
+4(1+\tilde G_{1\bar 1}+\tilde G_{2\bar 2} +\tilde G_{1\bar 1}\tilde
G_{2\bar 2}-\tilde G_{1\bar 2}\tilde G_{2\bar 1})=0\,.
\label{tildeG}\ee 
(Note that here we used the fact that we had chosen the complex
coordinates for the initial hyperk\"ahler base such that the volume
factor is unit, {\it i.e.} $V=1$.)

   It is clear that there always exist such $\omega^1$ and $\omega^2$
that satisfy (\ref{holotrans}).  For example, we can take $z^1=\int
H_1(\omega^i) d\omega^1$ and $z^2=\omega^2$. Also note that 
\bea
 A&=&-\frac{\rm
i}{4}(\tilde\partial-\bar{\tilde\partial})\partial_y G(y,\omega^i,\bar
\omega^i)=-\frac{\rm
i}{4}(\tilde\partial-\bar{\tilde\partial})\partial_y
\left(G(y,\omega^i,\bar \omega^i)+K_0(\omega^i,\bar \omega^i)\right)
\cr&=&-\frac{\rm
i}{4}(\tilde\partial-\bar{\tilde\partial})\partial_y \left(\tilde
G(y,z^i,\bar z^i)+K_0(z^i,\bar z^i)\right) =-\frac{\rm
i}{4}(\tilde\partial-\bar{\tilde\partial})\partial_y \tilde
G(y,z^i,\bar z^i)\,. \eea 
Thus the above transformation is a gauge transformation that
preserves our initial ansatz.  Hence we can set $g=1$ by this gauge
transformation.  It follows from (\ref{general3}) that $\kappa_0=0$.
Now, we have demonstrated that the system with $\kappa_1=0$ is
determined solely by the following basic equation for $G$ 
\be
\partial^2_y G
+4(1+G_{1\bar 1}+G_{2\bar 2}+G_{1\bar 1}G_{2\bar 2}-G_{1\bar
2}G_{2\bar 1})=0\,.\label{basicG2} \ee
It can be regarded as the $\kappa_1\rightarrow0$ limit of
(\ref{basicG1}) if we recover the $\kappa_1$ therein.  (In the
special case when the base space is the flat $\R^4$, the equation
(\ref{basicG2}) was also obtained in \cite{sant}, but with a
numerical error.  The factor ``8'' in equation (2.51) of \cite{sant}
should be ``16'' instead.)

     To summarize, we find that the Calabi-Yau metrics depend
on a discrete choice of the $\kappa$ function.  One is that
$\kappa=\alpha$, in which case the solution is completely determined
by one basic equation for $G$, given by (\ref{basicG1}). The other
is that $\kappa =0$, in which case the solution is completely
determined by the basic equation (\ref{basicG2}).

\section{The $\R^4$ base}

     Having obtained the general formalism for constructing the CY3
metrics from any hyperk\"ahler metric in four dimensions, we
consider explicit examples in this and the next sections. Note that all the hyperk\"ahler bases are related by a modification of K\"ahler potential. Therefore, they are equivalent to each other in our previously general construction. However, since the general basic equation is impossible to solve fully, different choices of hyperk\"ahler base will give different result when we construct the explicit metric in certain simplified ansatz.  The
simplest hyperk\"aler space is the Euclidean space $\R^4$.
An obvious choice is to use the complex coordinates $(z_1,z_2)$
directly, and the corresponding complex vielbein is given by 
\be\label{flatV1}
\tilde\epsilon^1=dz_1\,,~~~~\tilde\epsilon^2=dz_2\,. \ee 
Alternatively, one can write the $\R^4$ in terms of the
spherical-polar coordinates, {\it i.e.}
\be ds^2 = dr^2  + \ft14  r^2 (\sigma_1^2 + \sigma_2^2+\sigma_3^2)
\,, \ee
with the following choice of the complex vielbein
\be \tilde\epsilon^1 =\tilde e^1 + {\rm i}\, \tilde e^2=dr +
\fft{\rm i}2\, r\,\sigma_3\,,\qquad \tilde\epsilon^2=\tilde e^3 +
{\rm i}\, \tilde e^4=\fft12\,r\,(\sigma_1 + {\rm i}\,\sigma_2). \ee
Here, we define the $SU(2)$ Maurer-Cartan forms by
\bea \sigma_1 &=&\sin\psi\, d\theta - \sin\theta\, \cos\psi\,
d\phi\,,\cr \sigma_2&=&-\cos\psi\, d\theta -\sin\theta\,\sin\psi\,
d\phi\,,\cr \sigma_3&=& d\psi + \cos\theta\, d\phi\,. \eea
From the relation 
\be\label{flatV2} z_1=r\,\cos\fft\theta2\,\exp(\fft{\i(\psi+\phi)}2)
\,,~~~~z_2=r\,\sin\fft\theta2\,\exp(\fft{\i(\psi-\phi)}2)\,, \ee 
one can show that 
\be \bar{\tilde\epsilon}_a^{\mu}\partial_{\mu} z_i=0\,. \ee 
Therefore, the two choices of the complex vielbein (\ref{flatV1})
and (\ref{flatV2}) are compatible.

For our purpose, we find that the vielbein (\ref{flatV2}) is more
useful for simplifying equations under the isometry group of the
$S^3$ level surfaces. Under this choice, the general ansatz is
\bea ds^2 &=& (1+G_{1\bar1}) ({dr^2} + \ft14 \, r^2 \sigma_3^2) +
\ft14 (1+G_{2\bar2}) r^2 (\sigma_1^2 + \sigma_2^2)
\cr&&+(G_{1\bar2}+G_{2\bar1})(\ft12r\,dr\,\sigma_1+
\ft14r^2\,\sigma_3\,\sigma_2) -{\rm
i}(G_{1\bar2}-G_{2\bar1})(\ft12r\,dr\,\sigma_2-\ft14r^2\,\sigma_3
\,\sigma_1)\cr&&+ h^2\, dy^2 + \fft{1}{h^2}\, (d\alpha+ A )^2\,.
\eea
The inverse complex vielbein is given by
\be \tilde\epsilon_1 = \fft12(\tilde E_1 - {\rm i}\, \tilde
E_2)\,,\qquad \tilde\epsilon_2=\fft12(\tilde E_3 - {\rm i}\,\tilde
E_4)\,, \ee
where
\bea \tilde E_1&=&\;E_r\,,\cr \tilde E_2&=&\frac{2}{r}\;E_\psi,\cr
\tilde
E_3&=&\fft2r(\sin\psi\;E_\theta-\frac{\cos\psi}{\sin\theta}
E_{\phi}+\frac{\cos\theta\cos\psi}{\sin\theta}E_{\psi})\,,\cr
\tilde
E_4&=&\fft2r(-\cos\psi\;E_\theta-\frac{\sin\psi}{\sin\theta}E_{\phi}
+\frac{\cos\theta\sin\psi}{\sin\theta}E_{\psi})\,.
\eea
Note that $\Gamma^{\bar k}_{i\,\bar j}=\Gamma^{\bar k}_{i\,j}=0$ on
the K\"ahler manifold, thus we have
\bea\label{Gab} G_{a \bar b}
&=&
{\tilde\epsilon}_a^i\bar{\tilde\epsilon}_b^j
\partial_{z^i}\partial_{\bar z^j}G
= {\tilde\epsilon}_a^i\bar{\tilde\epsilon}_b^j
\nabla_{z^i}\nabla_{\bar z^j}G \cr&=&
{\tilde\epsilon}_a^{\mu}\bar{\tilde\epsilon}_b^{\nu}
\nabla_{\mu}\nabla_{\nu}G={\tilde\epsilon}_a^{\mu}
\bar{\tilde\epsilon}_b^{\nu}(\partial_{\mu}\partial_{\nu}
-\Gamma_{\mu\nu}^\lambda\partial_{\lambda})G\;. \eea
The explicit form of the $G_{a\bar b}$ is presented in (\ref{Gab1}).
The 1-form connection $A$ is given by 
\bea A&=&-\frac{\rm
i}{4}(\tilde\partial-\bar{\tilde\partial})\partial_y
G=-\frac{\rm
i}{4}(\tilde\epsilon^a\tilde\epsilon_a^{\mu}-\bar{\tilde\epsilon}^a
\bar{\tilde\epsilon}_a^{\mu})\partial_{\mu}\partial_yG
~~~~~~~~~~~~~~~~~~~~~~~~~~~~~~~~~~~~~~~~ \\
&=& \frac{\partial_r
\partial_yG}{8}\,r\,\sigma_3
-\fft{\partial_{\theta}\partial_yG}{4}\,\sin\theta\,d\phi
+\frac{\partial_{\phi}\partial_yG}{4\sin\theta}\,d\theta
-\frac{\partial_{\psi}
\partial_yG}{2r}dr-\frac{\partial_{\psi}\partial_yG}{4\sin\theta}\,
\cos\theta\,d\theta\,. \nn \eea 

Since the general equations are rather complicated, we shall further
suppose that the functions $G$ and $h$ depend on the coordinates
$(r,y)$ only as in \cite{sant}.  In such a radial ansatz, the resulting
metric has the $SU(2)\times U(1)^2$ isometry.  Note that we have
$G_{1\bar2}=0$ when $G=G(r,y)$. Thus the metric ansatz is reduced
to the following form
\be\label{flat} ds^2 = f_1 ({dr^2} + \ft14 \, r^2 \sigma_3^2) +
\ft14 f_2 r^2 (\sigma_1^2 + \sigma_2^2) + \fft{f_1\,f_2}{g^2}\, dy^2
+ \fft{g^2}{f_1\,f_2}\, (d\alpha + f_3 \sigma_3)^2\,. \ee
where
\bea &&f_2 \equiv1+\frac{1}{2r}\;\partial_{r}G\,,~~~g\equiv
f\,h^{-1}=\sqrt{f_1\,f_2}\,h^{-1}\,, \cr&& f_1 \equiv 1+\fft
14\;\partial_r^2G+\frac{1}{4r}\;\partial_{r}G=\fft{1}{2r} \del_r
(r^2 f_2)\,, \cr&& f_3 \equiv \frac{r}{8}\,\partial_r
\partial_yG=\ft14 \del_y (r^2 f_2)\,. \eea

\subsection{Case I: $\kappa_1=1$}
In this case, we have
\be g=\exp\left(-\fft{1}4\,\partial_yG\right)\,, \ee
and the system is determined solely by the basic equation
\be\label{flatcase1}
\partial_y\left[\exp\left(-\fft{1}2\,\partial_yG\right)\right]
=\fft{1}{2\,r^3}\partial_r\left[r^4\left(1+\fft1{2r}\partial_r
G\right)^2\right]\,. \ee
It is likely difficult to solve this equation fully. We obtain
two special solutions: one is just a direct product of Eguchi-Hanson
instanton and $\R^2$; the other is given by
\be ds^2=\fft{dy^2}{W} + \ft14 W y^2 (d\alpha - r^2 \sigma_3)^2 +
y^2 \Big(\fft{dr^2}{V} + \ft14 V r^2 \sigma_3^2 + \ft14 r^2(
\sigma_1^2 + \sigma_2^2)\Big)\,, \label{special1}\ee 
where
\be W= 1 - \fft{a}{y^6}\,,\qquad V=1 - r^2 - \fft{b}{r^4}\,.
\label{wv}\ee
The detailed derivation can be found in appendix B.1.
The metric (\ref{special1}) with $b=0$ was known in \cite{pp},
describing a higher dimensional generalization of Eguchi-Hanson instanton,
with $\R^2\times \CP^2$ topology and an asymptotic $\R^6/Z_3$. For $a=0$,
the metric is a cone of $Y^{p,q}$. The general solution describes a
resolution of the $Y^{p,q}$ cone, and the detailed global analysis can
be found in \cite{ooya,lpres,cclpv}.

      It should be emphasized that we have obtained only a special
solution to the basic equation (\ref{flatcase1}).  It would be
interesting to find the general solutions and examine the
corresponding metrics.

\subsection{Case II: $\kappa_1=0$}

In this case, we may use $f_2$ instead of $G$ as a basic function.
Then the basic equation (\ref{basicG2}) becomes
\bea\label{flatcase2}   \partial^2_yf_2+
\fft1{2r}\,\partial_r\left(\fft{1}{r^3}\partial_r(r^4f_2^2)\right)=0\,.
\eea
(This case was discussed in \cite{sant}.  However, there is an error
in the equation (2.52) for $G$.  The constant factor ``8'' should be
``16'' instead. This error propagates to the later metric results.)

\subsubsection{Some special solutions}

One way to solve (\ref{flatcase2}) is to consider the following
ansatz
\be f_2=  u(y)\, r^2 +  \xi(y)\,, \ee
then we find
\be f_1=2\;u\;r^2+\;\xi\,,~~~~~~f_3=\fft14(u_y\;r^2+\xi_y)\;r^2\,.
\ee
The functions $u$ and $\xi$ satisfy
\bea u_{yy}=-16\;u^2\,,\qquad\xi_{yy}=-12\;u\;\xi .\eea
An immediate solution is the degenerate case 
\be u(y)=0\,,\qquad \xi(y)=c_1\,y +c_2\,. \ee 
Beside this case, note that \bea
(u_{y})^2=-\frac{32}{3}(u^3-c^3)\,,\eea
\be dy^2 = \fft{3 du^2}{32(c^3 -  u^3)}\,, \ee
where $c$ is an integration constant. It would be better if we use
$u$ instead of $y$ as the coordinate. This implies that $\xi$
satisfies
\be 8 \,\left(c^3-u^3\right)\, \xi_{uu}-12 \,u^2 \,\xi_u +9 \,u
\,\xi =0 \,. \ee
The exact solution for $\xi(u)$ is
\bea \xi(u)&=&{}_2F_1\left(\frac{1}{12}
\left(1-\sqrt{19}\right),\frac{1}{12}
\left(1+\sqrt{19}\right);\frac{2}{3};\frac{u^3}{c^3}\right)C_1
\cr&&+ \; {}_2F_1\left(\frac{1}{12}
\left(5-\sqrt{19}\right),\frac{1}{12}
\left(5+\sqrt{19}\right);\frac{4}{3};\frac{u^3}{c^3}\right)u\;C_2
\eea
where ${}_2F_1$ is the hypergeometric function. When $c=0$, the
solution takes the simple form 
\bea
\xi(u)&=&|u|^{-\frac{1}{4}-\frac{\sqrt{19}}{4}}C_1+
|u|^{-\frac{1}{4}+\frac{\sqrt{19}}{4}}C_2\;.
\eea
Now the metric becomes 
\bea
ds^2_6&=&(2\;u\;r^2+\;\xi)\;(\;dr^2+\fft14r^2\;\sigma_3^2)
+\fft14(u\;r^2+\;\xi)\;r^2\;(\sigma_1^2+\sigma_2^2)
\cr&&+\frac{3(2\;u\;r^2+\;\xi)(u\;r^2+\;\xi)}{32(c^3-u^3)}\;du^2
\cr&&+\frac{1}{(2\;u\;r^2+\;\xi)(u\;r^2+\;\xi)}\;
\bigg(d\alpha+{\sqrt{\fft23(c^3-u^3)}}\;
(r^2+\xi_u)\;r^2\;\sigma_3\bigg)^2\;.
\eea 
We must have $u<c$ to keep the metric real. If $u>0$, the range of
$r$ is $(0,\infty)$. If $u<0$, then we must have $\xi>0$, and the
range of $r$ is constrained. Especially, when $c\leq0$, the
$r=\infty$ region is not reachable. The metrics have no asymptotic
cone over Einstein-Sasaki spaces and they develop a power-law
curvature singularity when $f_1f_2=0$.  This is rather different
from the case of $\kappa_1=1$, where the non-vanishing $g$ in the
metric (\ref{flat}) allows non-singular collapsing of the cycles.

\subsubsection{Separation of variables}

  We can also solve the equation (\ref{flatcase2}) by separation of
variables, namely $f_2=u(y) \zeta(r)$.  Substituting this ansatz to
(\ref{flatcase2}), we have
\be u_{yy} + \ft32 k\, u^2=0\,,\qquad \fft{1}{r}
\partial_r(r^{-3} \partial_r (r^4 \zeta^2)) +3 k\, \zeta=0\,,
\ee
where $k$ is an arbitrary constant. The first equation implies that
\be dy^2 = \fft{du^2}{k(u^3-c^3)}\,. \ee
The solution for the second equation clearly exists, although there
appears to have no explicit analytical form, except for the case
with $k=0$, for which, the solution for $f_2$ is given by
(\ref{f2solx}).

\subsubsection{The formal general solution}

Expanding $f_2$ by Taylor series of $y$
\be f_2(r,y)=\sum_{n=0}^{\infty} u_n(r)y^n\,,\ee 
Substituting this into (\ref{flatcase2}), we find the recursion
relation 
\bea
\frac{1}{2r}\partial_r\left(\frac{1}{r^3}{\partial_r}
(\sum_{q=0}^{n}u_qu_{n-q}\,r^4)\right)+(n+2)(n+1)u_{n+2}=0
\,. \eea 
Given the two arbitrarily functions $u_0(r)$ and $u_1(r)$, we can
determine all the $u_n$ for $n\ge2$.  The general solution of our
system can thus be written formally by the Taylor series of $y$. If
we restrict that both $u_0$ and $u_1$ are in the form
$u\,r^2+\xi_0$, we find that all the $u_n$'s are in the form
$u\,r^2+\xi$. It is consistent with the solution we obtained
previously.

If we require that there be a maximum $n_{max}=N$ for no vanishing
$u_n$, we shall obtain polynomial solutions on $y$. There will be
$2N+1$ constraint equations for an $N$-th order polynomial solution
in general except for $N=0$.

\noindent{\bf $N=0$:} The only equation in this case is 
\bea
\frac{1}{2r}\partial_r\left(\frac{1}{r^3}{\partial_r}
(u_0^2\,r^4)\right)=0
\,. \eea 
We find that the solution is 
\bea f_2=\sqrt{1-\fft{a^4}{r^4}}\,b_0 \eea 
The corresponding metric is nothing but a direct product of the
Eughchi-Hanson instanton
and the $\R^2$.

\noindent{\bf $N=1$:} The equations in this case are 
\be\frac{1}{2r}\partial_r\left(\frac{1}{r^3}{\partial_r}
(u_0^2\,r^4)\right)=0 \,,\quad
\frac{1}{r}\partial_r\left(\frac{1}{r^3}{\partial_r}
(u_0u_{1}\,r^4)\right)=0\,,\quad
\frac{1}{2r}\partial_r\left(\frac{1}{r^3}{\partial_r}
(u_1^2\,r^4)\right)=0\,. \ee 
the solution is 
\bea f_2=\sqrt{1-\fft{a^4}{r^4}}\,(b_0+b_1\, y)\,.\label{f2solx}\eea

\noindent{\bf $N=2$:} The equations in this case are 
\bea &&\frac{1}{2r}\partial_r\left(\frac{1}{r^3}{\partial_r}
(u_0^2\,r^4)\right)+2u_{2}=0\,,\qquad
\frac{1}{r}\partial_r\left(\frac{1}{r^3}{\partial_r}
(u_0u_1\,r^4)\right)=0\,,\cr
&&\frac{1}{2r}\partial_r\left(\frac{1}{r^3}{\partial_r}
\left((2u_0u_2+u_1^2)\,r^4\right)\right)=0 \,,\qquad
\frac{1}{r}\partial_r\left(\frac{1}{r^3}{\partial_r}
(u_1u_2\,r^4)\right)=0\,,\cr
&&\frac{1}{2r}\partial_r\left(\frac{1}{r^3}{\partial_r}
(u_2^2\,r^4)\right)=0 \,. \eea 
We find that there is no consistent solution since the solution for
the last three equations is in contradiction with the first
equation. Similarly, it can be shown that there is no consistent
solutions for all finite $N\geq2$ since the solution for the last
$N+1$ equations are in contradiction with the $(N-1)$'th equation.

\section{The triaxial base}

In this section, we consider the triaxial hyperk\"ahler metric with
the $SU(2)$ isometry.  The metric is of the form
\be ds_4^2 =d\rho^2 + a_1^2\, \sigma_1^2 + a_2^2\,\sigma_2^2+a_3^2\,
\sigma_3^2\, , \ee
where $a_i$'s are functions of $\rho$ only, and they satisfy the
first order equations
\be \dot a_1=\frac{a_2^2+a_3^2-a_1^2}{2a_2a_3}\,,\qquad \dot
a_2=\frac{a_1^2+a_3^2-a_2^2}{2a_1a_3}\,,\qquad \dot
a_3=\frac{a_1^2+a_2^2-a_3^2}{2a_1a_2}\,. \ee
where a dot denotes a derivative with respect to $\rho$. We find
that the most general solution can be written as follows
\be ds^2_4 = \sqrt{\frac{1}{W\,\tilde W}}dr^2 + \fft14
r^2\,(\sqrt{{W\,\tilde W}}\, \sigma_1^2+\sqrt{\fft{\tilde W}W}
\sigma_2^2+\sqrt{\fft W{\tilde W}}\sigma_3^2)\,,\label{triaxialeh}
\ee
where
\be W=1 - \fft{a^4}{r^4}\,,\qquad \tilde W = 1 - \fft{b^4}{r^4}\,.
\label{wtildew} \ee
It is of interest to note that one can also introduce a cosmological
constant and construct the triaxial Einstein-K\"ahler spaces. The
three first-order equations were obtained in \cite{triaxialek}. Only
two exact solutions were known: one describes an $\CP^2$
\cite{bougib} and the other, a direct product $S^2\times S^2$
\cite{pope,com7}. The metrics are
\bea ds_{\CP^2}^2 &=& d\rho^2 + \sin^2\rho\, \sigma_1^2 +
\cos^2\rho\, \sigma_2^2 + \cos^22\rho\, \sigma_3^2\,,\cr
ds_{S^2\times S^2}^2 &=& d\rho^2 + \sin^2\rho\, \sigma_1^2 +
\sigma_2^2 + \cos^2\rho\, \sigma_3^2\,,\label{cp2s2s2} \eea

For the generic choice of constant parameters $a$ and $b$, the
metric (\ref{triaxialeh}) contains a naked power-law singularity at
$r=\max(a,b)$. When $b=a$, the metric reduces to the Eguchi-Hanson
instanton, given by
\be ds_4^2 =\fft{dr^2}{W} + \ft14 W\, r^2 \sigma_3^2 + \ft14
 r^2 (\sigma_1^2 + \sigma_2^2)\,. \ee
We shall first consider this biaxial hyperk\"ahler base. One way to
choose the complex vielbein is
\be\label{eh1} \tilde\epsilon^1 = W^{-\fft12}\,dr + \fft{\rm i}2\,
W^{\fft12}\,r\,\sigma_3\,,\qquad
\tilde\epsilon^2=\fft12\,r\,(\sigma_1 + {\rm i}\,\sigma_2). \ee
It is easy to see that such a choice will lead to the same radial
ansatz as in the $\R^4$ case. Fortunately, there are three possible
K\"ahler structures for a hyperk\"ahler space. For the flat base,
all the three choices give same ansatz. For the
Eguchi-Hanson base, the remaining two choices are equivalent to each
other but inequivalent to (\ref{eh1}). The corresponding complex
vielbein is given by
\be\label{eh2} \tilde\epsilon^1 =\tilde e^1 + {\rm i}\, \tilde e^2 =
W^{-\fft12}\,dr + \fft{\rm i}2\,r\, \sigma_3 \,,\qquad
\tilde\epsilon^2=\tilde e^3+ {\rm i}\, \tilde
e^4=\fft12\,r\,(W^{\fft12}\,\sigma_1 + {\rm i}\,\sigma_2), \ee where
we have permuted the $\sigma_i$'s for convenience. Correspondingly,
the inverse complex vielbein is given by
\be \tilde\epsilon_1 = \fft12(\tilde E_1 - {\rm i}\, \tilde
E_2)\,,\qquad \tilde\epsilon_2=\fft12(\tilde E_3 - {\rm i}\,\tilde
E_4)\,, \ee
where
\bea \tilde E_1&=&W^{\fft12}\;E_r\,,\cr \tilde
E_2&=&\frac{2}r\;E_\psi,\cr \tilde
E_3&=&\fft2{r\,W^{\fft12}}(\sin\psi\;E_\theta-
\frac{\cos\psi}{\sin\theta}E_{\phi}+
\frac{\cos\theta\cos\psi}{\sin\theta}E_{\psi})\,,\cr \tilde
E_4&=&\fft2r(-\cos\psi\;E_\theta-\frac{\sin\psi}{\sin\theta}
E_{\phi}+\frac{\cos\theta\sin\psi}{\sin\theta}E_{\psi})\,.
\eea

It follows from the relation (\ref{Gab}) that $G_{a\bar b}$ can
be obtained. The result is presented in (\ref{Gab2}).
The 1-form $A$ is given by 
\bea A&=&-\frac{\rm
i}{4}(\tilde\partial-\bar{\tilde\partial})\partial_y
G =-\frac{\rm
i}{4}(\tilde\epsilon^a\tilde\epsilon_a^{\mu}-
\bar{\tilde\epsilon}^a\bar{\tilde\epsilon}_a^{\mu})
\partial_{\mu}\partial_yG  \cr&=& \frac{\partial_r
\partial_yG}{8}\,r\,W^{\fft12}\,\sigma_3 +\frac{\partial_\theta
\partial_yG}{4}(W^{\fft12}\,\cos\psi\,\sigma_1+
W^{-\fft12}\,\sin\psi\,\sigma_2)\cr && +\frac{\partial_\phi
\partial_yG}{4\,\sin\theta}(W^{\fft12}\,\sin\psi\,\sigma_1-
W^{-\fft12}\,\cos\psi\,\sigma_2) \cr&&-\frac{\partial_{\psi}
\partial_yG}{4} \left(\fft2{r\,W^{\fft12}}\,dr+W^{\fft12}\,
\cot\theta\,\sin\psi\,\sigma_1-W^{-\fft12}\,\cot\theta\,\cos\psi\,
\sigma_2\right) \,.\eea 

The metric in the radial ansatz is then given by the following form
\be\label{eh} ds^2 = f_1 (\frac{dr^2}W+ \ft14 \, r^2 \sigma_3^2) +
\ft14 f_2 r^2 (W\sigma_1^2 + \sigma_2^2) + \fft{f_1\,f_2}{g^2}\,
dy^2 + \fft{g^2}{f_1\,f_2}\, (d\alpha + f_3 \sigma_3)^2\,. \ee
where
\bea &&f_2
=1+\frac{1}{2r}\;\partial_{r}G\,,~~~g=f\,h^{-1}=\sqrt{f_1\,f_2}\,h^{-1}\,,
\cr&& f_1 = 1+\fft
14W\;\partial_r^2G+\frac{1}{4r}\left(1+\frac{a^4}{r^4}\right)\;
\partial_{r}G=\fft{W^{\fft12}}{2r}\partial_r\left(r^2\,W^{\fft12}
\,f_2\right)\,,\cr&& f_3 =\frac{r\,W^{\fft12}}{8}\,\partial_r
\partial_yG=\ft14 r^2W^{\fft12}\,\del_y  f_2\,. \eea
The metric has the isometry of $SU(2)\times U(1)$.

\subsection{Case I: $\kappa_1=1$}

In this case, we have
\bea && g=\exp\left(-\fft{1}4\,\partial_yG\right)\,, \eea
and the system is determined solely by the basic equation
\bea\label{ehcase1} &&
\partial_y\left[\exp\left(-\fft{1}2\,\partial_yG\right)\right]
=\fft{1}{2\,r^3}\partial_r\left[(r^4-a^4)\left(1+\fft1{2r}
\partial_r G\right)^2\right]\,. \eea
We obtain some special solutions. One describes an $\R^2\times \CP^2$
instanton that is asymptotic to the $\R^6/Z_3$.  Another describes
an $\R^2\times S^2\times S^2$ instanton  that is asymptotic to the
cone over $T^{1,1}/Z_2$.  The details are presented in appendix B.2.

\subsection{Case II: $\kappa_1=0$}

In this case, we may use $f_2$ instead of $G$ as the basic function.
Then the basic equation (\ref{basicG2}) becomes
\bea\label{ehcase2}
\partial^2_yf_2+\fft1{2r}\,
\partial_r\left(\fft1{r^3}\partial_r\left((r^4-a^4)f_2^2\right)
\right)=0\,.
\eea

\subsubsection{Some special solution}

If $f_2$ depends
only on $y$, the solution is simply $f_2=c_1\,y+c_2$ and the metric
is
\bea ds^2 &=& (c_1\,y+c_2) (\fft{dr^2}{W} + \fft14 r^2\,\sigma_3^2)
+ \ft14 (c_1\,y+c_2) r^2 ( W\, \sigma_1^2 + \sigma_2^2) \cr&&+
(c_1\,y+c_2)^2\, dy^2 + \fft{1}{(c_1\,y+c_2)^2}\, (d\alpha +
\fft{c_1}{4} r^2\,W^{\fft12} \sigma_3)^2\,. \eea
If  $f_2= u(y)\, r^2 $, the basic function is simplified to 
\be u_{yy}=-16 u^2\,. \ee 
It implies
\be dy^2 = \fft{3 du^2}{32(c^3 -  u^3)}\,. \ee
Then we can use $u$ instead of $y$ as the coordinate, and the metric
is given by 
\bea ds^2_6&=&(1+W)\,u\, r^2 \,(\fft{dr^2}{W} + \fft14
r^2\,\sigma_3^2) + \ft14 u\, r^4 ( W\, \sigma_1^2 + \sigma_2^2)
\cr&&+\frac{3(1+W)\,u^2\,r^4}{32(c^3-u^3)}\;du^2
+\frac{1}{(1+W)\,u^2\,r^4}\;\bigg(d\alpha+{\sqrt{\fft23(c^3-u^3)}}
\;r^4\,W^{\fft12}\;\sigma_3\bigg)^2\;.
\eea 

\subsubsection{The formal general solution}

The general solution can be expressed formally by Taylor series of
$y$ as previously. Expanding $f_2$ as 
\be f_2(r,y)=\sum_{n=0}^{\infty} u_n(r)y^n\,,\ee 
we find the recursion relations 
\bea
\frac{1}{2r}\partial_r\left(\frac{1}{r^3}{\partial_r}
\left(\sum_{q=0}^{n}u_qu_{n-q}\,(r^4-a^4)\right)\right)
+(n+2)(n+1)u_{n+2}=0
\,. \eea 
Given the two arbitrary functions $u_0(r)$ and $u_1(r)$, we can
determine all the $u_n(n>1)$'s by the above recursion relations.
Then the general solution of our system can be written formally by
the Taylor series of $y$. If we restrict that both $u_0$ and $u_1$
are proportion to $r^2$, we find that all the $u_n$'s are proportion
to $r^2$ by the recursion relations. It is consistent with the
previous result. If we require there is a maximum $n_{max}=N$ for no
vanishing $u_n$, we shall obtain polynomial solutions on $y$. There
are $2N+1$ constraint equations for an $N$-th order polynomial
solution. We now examine these equations as follows.
\\ {\bf $N=0$:} The only equation in this case is
\bea
\frac{1}{2r}\partial_r\left(\frac{1}{r^3}{\partial_r}
\left(u_0^2\,(r^4-a^4)\right)\right)=0
\,. \eea 
We find that the solution is 
\bea f_2=\lambda\sqrt{\fft{r^4-b^4}{r^4-a^4}}\,. \eea 
It gives rise to a direct product of the triaxial metric
(\ref{triaxialeh}) and the $\R^2$.
\\ {\bf $N=1$:} The equations in this case are
\bea &&\frac{1}{2r}\partial_r\left(\frac{1}{r^3}{\partial_r}
\left(u_0^2\,(r^4-a^4)\right)\right)=0 \,,\qquad
\frac{1}{r}\partial_r\left(\frac{1}{r^3}{\partial_r}
\left(u_0u_{1}\,(r^4-a^4)\right)\right)=0\,,\cr &&
\frac{1}{2r}\partial_r\left(\frac{1}{r^3}{\partial_r}
\left(u_1^2\,(r^4-a^4)\right)\right)=0 \,. \eea 
The solution is 
\bea f_2=\sqrt{\fft{r^4-b^4}{r^4-a^4}}\,(\lambda_0+\lambda_1\, y)
\eea 
As in the flat base case, we find again that there are no finite
polynomial solutions of the $y$ coordinate, for any order $N\ge 2$.

\section{Conclusions}

    In this paper, we examine the metric construction for Calabi-Yau
3-folds proposed in \cite{sant}.  The essence of the construction is
to add a complex line bundle over a four dimensional hyperk\"ahler
structure, with a simple deformation where the four-dimensional
K\"ahler potential is modified.  The resulting metric ansatz has at
least one Killing direction.  It was demonstrated for the $\R^4$
base that the condition for the CY3 metrics could be indeed
satisfied.

    We extend the construction and obtain the general formalism for a
generic hyperk\"ahler base.  Furthermore, we find that the ansatz
for the holomorphic $(3,0)$-form should be generalized to have a
$U(1)$ factor.  This allows us to construct the CY3 metrics that are
asymptotic to the cones of Einstein-Sasaki spaces.  There can be a
discrete choice for the $U(1)$ factor.  One is that it depends on
the fibre $U(1)$ coordinate only, and consequently the equations of
motion are reduced to one differential equation on the modified
K\"ahler potential.  The other is that the $U(1)$ factor vanishes.
In this case, the metrics are determined by a differential equation
that is the singular limit of the previous one.

We then construct explicit metrics with two examples of the
hyperk\"ahler bases. One is the $\R^4$, and the other is the
triaxial metric with $SU(2)$ isometries. In both cases, we obtain
explicit cohomogeneity-2 metrics.  With the $U(1)$ factor, we obtain
a general class of solutions that describe a resolution of the cone
over $Y^{p,q}$ spaces.  For the case with vanishing $U(1)$ factor,
we obtain singular metrics with no asymptotically conical region.
The solutions are governed by two arbitrary functions of the radial
variable of the hyperk\"ahler spaces.

    The general construction we have obtained allows one to construct
a wide class of CY3 metrics with at least one Killing direction.  It
is of great interest to investigate whether new complete metrics on
the non-compact manifolds can arise.

\appendix

\section{Explicit $G_{a\bar b}$}

In this appendix we give the explicit expressions for the $G_{a\bar b}$
defined by (\ref{Gab}).  For the $\R^4$ base discussed in section 3,
we find that they are given by
\bea G_{1\bar1}&=&\fft
14\;\partial_r^2G+\frac{1}{r^2}\;\partial_{\psi}^2G
+\frac{1}{4r}\;\partial_{r}G\,,\cr
G_{2\bar2}&=&\frac{1}{r^2}\;\partial_{\theta}^2G+
\frac{1}{r^2\sin^2\theta}\;\partial_{\phi}^2G
-\frac{2\cos\theta}{r^2\sin^2\theta}\;\partial_{\phi}\partial_{\psi}G
+\frac{\cos^2\theta}{r^2\sin^2\theta}\;\partial_{\psi}^2G\cr &&
+\frac{1}{2r}\;\partial_{r}G
+\frac{\cos\theta}{r^2\sin\theta}\;\partial_{\theta}G\,,\cr
G_{1\bar2} &=&e^{{\rm
i}\psi}\left[-\frac{1}{2r\sin\theta}\partial_{r}\partial_{\phi}G
+\frac{\cos\theta}{2r\sin\theta}\;\partial_{r}\partial_{\psi}G
-\frac{1}{r^2}\;\partial_{\theta}\partial_{\psi}G
\right.\cr&&\left.~~~~~~~+{\rm
i}\left(-\frac{1}{2r}\partial_r\partial_{\theta}G
+\frac{1}{r^2\sin\theta}\partial_{\phi}\partial_{\psi}G
-\frac{\cos\theta}{r^2\sin\theta}\partial^2_{\psi}G\right)\right]
\cr&=&(G_{2\bar1})^*\,.\label{Gab1} \eea 
For the triaxial base discussed in section 4, they are more complicated,
given by
\bea G_{1\bar1}&=&\fft14\left(1 -
\fft{a^4}{r^4}\right)\partial_r^2G+\fft{1}{r^2}
\partial_\psi^2G+\fft1{4r}\left(1
+ \fft{a^4}{r^4}\right)\partial_rG\,,\cr
G_{2\bar2}&=&\fft{r^4-a^4\,{\cos^2\psi}}{r^2(r^4-a^4)}
\partial_\theta^2G -\fft{a^4\,\sin2\psi}{r^2(r^4-a^4)\,
\sin\theta}\partial_\theta\partial_\phi G +\fft{a^4\,
\cos\theta\,\sin2\psi}{r^2(r^4-a^4)\,\sin\theta}
\partial_\theta\partial_\psi G\cr
&&+\fft{r^4-a^4\,{\sin^2\psi}}{r^2(r^4-a^4)\,\sin^2\theta}
\partial_\phi^2G -\fft{2\,(r^4-a^4\,{\sin^2\psi})
\cos\theta}{r^2(r^4-a^4)\,\sin^2\theta}
\partial_\phi\partial_\psi G \cr&&+\fft{(r^4-a^4\,{\sin^2\psi})
\cos^2\theta}{r^2(r^4-a^4)\,\sin^2\theta}\partial_\psi^2G
+\frac{1}{2r}\partial_r G
+\fft{(r^4-a^4\,{\sin^2\psi})\cos\theta}{r^2(r^4-a^4)
\,\sin\theta}\partial_\theta G
\cr&&+\fft{a^4\,{\cos\theta\,\sin2\psi}}{r^2(r^4-a^4)\,
\sin^2\theta}\partial_\phi G
-\fft{a^4\,{(1+\cos^2\theta)\sin2\psi}}{2\,r^2(r^4-a^4)\,
\sin^2\theta}\partial_\psi G\,,\cr
G_{1\bar2}&=&\fft{\sin\psi}{2\,r}\partial_r\partial_\theta
G-\fft{\cos\psi}{2\,r\,\sin\theta}\partial_r\partial_\phi G
+\fft{\cos\theta\,\cos\psi}{2\,r\,\sin\theta}
\partial_r\partial_\psi
G -\fft{\cos\psi}{r^2}\partial_\theta\partial_\psi G \cr &&
-\fft{\sin\psi}{r^2\,\sin\theta}\partial_\phi\partial_\psi G
+\fft{\cos\theta\,\sin\psi}{r^2\,\sin\theta} \partial_\psi^2 G
-\fft{a^4\,{\sin\psi}}{r^2(r^4-a^4)}\partial_\theta G\cr &&
+\fft{a^4\,{\cos\psi}}{r^2(r^4-a^4)\,\sin\theta}\partial_\phi G
-\fft{a^4\,\cos\theta\,{\cos\psi}}{r^2(r^4-a^4)\,\sin\theta}
\partial_\psi G \cr&&+\,{\rm
i}\left[-\fft{W^{\fft12}\cos\psi}{2r}\partial_r\partial_\theta G
-\fft{W^{\fft12}\sin\psi}{2r\,\sin\theta}\partial_r
\partial_\phi G +\fft{W^{\fft12}\cos\theta\,\sin\psi}{
2r\,\sin\theta}\partial_r\partial_\psi G\right.\cr&&
\left.\qquad-\fft{\sin\psi}{r^2\,W^{\fft12}}\partial_\theta\partial_\psi
G +\fft{\cos\psi}{r^2\,
W^{\fft12}\,\sin\theta}\partial_\phi\partial_\psi G
-\fft{\cos\theta\,\cos\psi}{r^2\,W^{\fft12}\,\sin\theta}
\partial_\psi^2 G -\fft{a^4\,{\cos\psi}}{r^6\,W^{\fft12}}
\partial_\theta G \right.\cr&&\left.~~~~~~-\fft{a^4\,
{\sin\psi}}{r^6\,W^{\fft12}\,\sin\theta}\partial_\phi G
+\fft{a^4\,\cos\theta\,{\sin\psi}}{r^6\,W^{\fft12}\,
\sin\theta}\partial_\psi G\right]=(G_{2\bar1})^*\,.\label{Gab2}\eea 

\section{Detailed derivation for the $\kappa=1$ solutions}

\subsection{The $\R^4$ base}

For the $\kappa=1$ case, the system is reduced to one basic equation,
given by (\ref{flatcase1}). In this appendix, we obtain
some special solutions by considering the following ansatz
\be g^2=u_1(r)\,(a_3\,y^3+a_2\,y^2+3 a_1^2\,y+\tilde a_0)\,. \ee
Correspondingly, we have
\bea &&-\fft{1}2\,G=(2\log
g-3)\,y-\sum_{i=1}^3y_i\log(y-y_i)+u_2(r) \cr&=&y\log
u_1+y\log(a_3y^3+a_2y^2+3a_1^2y+\tilde
a_0)-3y-\sum_{i=1}^3y_i\log(y-y_i)+u_2(r)\,, \eea
where $y_i$'s are the roots of the equation
$a_3\,y^3+a_2\,y^2+3a_1^2\,y+\tilde a_0=0$.  Substituting this into
the the basic equation (\ref{flatcase1}), we have
\bea\label{flatrpoly}
\fft{1}{2\,r^3}\partial_r\left[r^2\,\left(\fft{u'_1}{u_1}\right)^2\right]
&=&3a_3\,u_1\,, \cr
\fft{1}{2r^3}\partial_r\left[r^2\,\left(u'_2-\,r\right)
\fft{u'_1}{u_1}\right]&=&a_2\,u_1\,,\cr
\fft{1}{2\,r^3}\partial_r\left[r^2\,
\left(u'_2-\,r\right)^2\right]&=&3a_1^2\,u_1\,.
\eea

If $a_3=0$, the consistency requires also $a_2=0$. Then we get
\bea u_1={\rm constant} \,,~~~~~f_2=1-\fft{u_2'}r=a_1
\sqrt{\fft{3u_1}{2}\left(1+\fft{a}{r^4}\right)}\,. \eea
After absorbing the redundant parameter by rescaling, the
metric is given by
\be ds^2 = \fft1{\sqrt{1+\fft{a^4}{r^4}}} ({dr^2} + \ft14 \, r^2
\sigma_3^2) + \fft{r^2}4 \sqrt{1+\fft{a^4}{r^4}} (\sigma_1^2 +
\sigma_2^2) +  dy^2 +  d\alpha^2\,. \ee
Further making the coordinate transformation 
\be (r^4+a^4)^{\fft14}\rightarrow r,\ee 
we get
\be ds^2 = \fft{dr^2}{1 - \fft{a^4}{r^4}} + \fft{r^2}{4}\, (1 -
\fft{a^4}{r^4})\, \sigma_3^2 + \ft14
 r^2 (\sigma_1^2 + \sigma_2^2) +  dy^2 +  d\alpha^2\,. \ee
This is nothing but a direct product of the Eughchi-Hanson instanton
and the $\R^2$.

If $a_3\neq0$, we can always set $a_3=1$ by
the redefinition of $u$. Then the consistency of the equations
implies
\bea u_2=\fft{1}{2}\,r^2+ a_1\log{u_1}\,,~~~~~~a_2=3\,a_1\,.
\eea
Thus
\bea g^2&=&u_1(r)\,(y^3+3a_1\,y^2+3 a_1^2\,y+\tilde
a_0)=u_1(r)\,[(y+a_1)^3+a_0]\,, \cr
r^2\,f_2&=&r^2\,(1+\frac{1}{2r}\;\partial_{r}G)= -(y+
a_1){r}\partial_r\log  u_1\,, \cr r^2\,f_1 &=& \fft{r}{2} \del_r
(r^2 f_2)=-(y+ a_1)\fft{r}{2}\del_r(r\,\partial_r\log  u_1)\,, \cr
f_3 &=&\ft14 \del_y (r^2 f_2)=-\frac{r}{4}\partial_r\log  u_1\, \cr
h^2 &=&\fft{f_1f_2}{g^2}=\fft{(y+a_1)^2\,\del_r(r\,\partial_r\log
u_1)\, \partial_r\log  u_1}{2\,[(y+a_1)^3+a_0]\,r^2\,u_1}
=\fft{3\,(y+a_1)^2}{2\,[(y+a_1)^3+a_0]}. \eea
Obviously, we can set $a_1=0$ by a coordinate transformation. Then
the metric is given by
\bea ds^2 &=& -\frac{y}{2}\,\partial_r \rho\, (\fft{dr^2}{r}
+ \fft{1}4 \,r\, \sigma_3^2)  -\frac{y}{4}\rho\, (\sigma_1^2
+ \sigma_2^2) \cr&& + \fft{3\,y^2 }{2\,(y^3+a_0)}\, dy^2 +
\fft{2\,(y^3+a_0)}{3\,y^2 }\, (d\alpha
 -\frac{1}{4}\rho\, \sigma_3)^2\,,\eea
where
\bea \rho&=&r\,\partial_r\log  u_1. \eea
Taking $\rho$ instead of $r$ as the radial coordinate and supposing
\bea r\,\partial_r=V(\rho)\,\partial_\rho\,, \eea
the metric becomes
\bea\label{flatsol1} ds^2 &=& -\frac{y}{2}\,
(\fft{d\rho^2}{V} + \fft{V}4 \, \sigma_3^2)
-\frac{y}{4}\rho\, (\sigma_1^2 + \sigma_2^2) \cr&& +
\fft{3\,y^2 }{2\,(y^3+a_0)}\, dy^2 +
\fft{2\,(y^3+a_0)}{3\,y^2 }\, (d\alpha
 -\frac{1}{4}\rho\, \sigma_3)^2\,.\eea
The first equation in (\ref{flatrpoly}) becomes
\bea\label{flatbasic} &&3
\,u_1=\fft{V}{2\,r^4}\partial_\rho(\rho^2)=\fft{\rho\,V}{r^4}\,.
\eea
Thus
\bea \rho=V\,\partial_{\rho}\log u_1 =\fft
1{\rho}\partial_\rho{\left(\rho\,V\right)}-4\,. \eea
The solution is
\bea V=\left(\fft13\rho^3+2\rho^2+b_0\right)\rho^{-1}\,. \eea
Let $\rho\rightarrow-\rho$ for convenience, then the CY3 metric is
given by
\bea ds^2 &=& \frac{y}{2 }\,
(\fft{d\rho^2}{2\rho-\fft13\rho^2+\fft{b_0}{\rho}} +
\fft{2\rho-\fft13\rho^2+\fft{b_0}{\rho}}4 \, \sigma_3^2)
+\frac{y}{4 }\rho\, (\sigma_1^2 + \sigma_2^2) \cr&& +
\fft{3\,y^2 }{2 \,(y^3+a_0)}\, dy^2 +
\fft{2 \,(y^3+a_0)}{3\,y^2 }\, (d\alpha
 +\frac{1}{4}\rho\, \sigma_3)^2\,.\eea
After making some appropriate coordinate transformations, rescaling
of the metric and renaming the constants, the metric can be cast into
(\ref{special1}).

\subsection{The triaxial base}

The basic equation for the $\kappa=1$ solutions is given by (\ref{ehcase1}).
Since the structure is quite similar with that of the flat $\R^4$ base,
we take the same ansatz as in that case, namely
\bea  g^2&=&u_1(r)\,(a_3\,y^3+a_2\,y^2+3 a_1^2\,y+\tilde a_0)\,.
\eea 
Thus we have 
\bea &&-\fft{1}2\,G=(2\log
g-3)\,y-\sum_{i=1}^3y_i\log(y-y_i)+u_2(r) \cr&=&y\log
u_1+y\log(a_3y^3+a_2y^2+3a_1^2y+\tilde
a_0)-3y-\sum_{i=1}^3y_i\log(y-y_i)+u_2(r)\,, \eea
where $y_i$'s are the roots of the equation
$a_3\,y^3+a_2\,y^2+3a_1^2\,y+\tilde a_0=0$.  Then, the basic
equation implies
\bea\label{ehrpoly}
\fft{1}{2\,r^3}\partial_r\left[r^2\,W\,\left(\fft{u'_1}{u_1}
\right)^2\right]&=&3a_3\,u_1\,, \cr
\fft{1}{2r^3}\partial_r\left[r^2\,W\,\left(u'_2-
r\right)\fft{u'_1}{u_1}\right]&=& a_2\,u_1\,, \cr
\fft{1}{2\,r^3}\partial_r\left[r^2\,W\,\left(u'_2-r\right)^2
\right]&=&3a_1^2\,u_1\,. \eea

If $a_3=0$, the consistency requires also $a_2=0$. The nontrivial
solution is given by
\be u_1={\rm constant} \,,~~~~ f_2 =1-\fft{u_2'}r=
a_1\sqrt{\fft{3u_1}{2}}\sqrt{\frac{r^4-b^4}{r^4-a^4}}\,. \ee 
After absorbing the redundant parameter by rescaling, the corresponding
metric is
\bea ds^2 = \sqrt{\frac{1}{W\,\tilde W}}dr^2 + \fft14
r^2\,(\sqrt{{W\,\tilde W}}\, \sigma_1^2+\sqrt{\fft{\tilde W}W}
\sigma_2^2+\sqrt{\fft W{\tilde W}}\sigma_3^2) +  dy^2 + d\alpha^2\,.
\eea
This is just a direct product of the triaxial metric
(\ref{triaxialeh}) and the $\R^2$. It implies that the radial ansatz for the
CY3 metric based on the triaxial hyperk\"ahler base will be
equivalent to (\ref{eh}). Therefore, up to coordinate
transformations which permutate the three $\sigma_i$'s, there will
be no further radial ansatz coming from the triaxial base.

 For non-vanishing $a_3$, we can set it to unity by redefinition of $u$.
Then the consistency of the equations implies
\be u_2=\fft{1}{2}\,r^2+ a_1\log{u_1}\,,~~~~~~a_2=3\,a_1\,. \ee
Thus
\bea g^2&=&u_1(r)\,(y^3+3a_1\,y^2+3 a_1^2\,y+\tilde
a_0)=u_1(r)\,[(y+a_1)^3+a_0]\,, \cr
r^2\,f_2&=&r^2\,(1+\frac{1}{2r}\;\partial_{r}G)= -(y+
a_1){r}\partial_r\log  u_1\,, \cr r^2\,f_1 &=& \ft12
r\,W^{\fft12}\partial_r\left(r^2\,W^{\fft12}\,f_2\right)=-\ft12 (y+
a_1) r\,W^{\fft12}\del_r(r\,W^{\fft12}\,\partial_r\log u_1)\,, \cr
f_3 &=&\ft14 r^2W^{\fft12}\,\del_y f_2=-\ft14r\,W^{\fft12}
\partial_r\log  u_1\, \cr
h^2&=&\fft{f_1f_2}{g^2}=\fft{(y+a_1)^2\,W^{\fft12}\,\del_r(r\,W^{\fft12}
\,\partial_r\log u_1)\, \partial_r\log u_1}{2
\,[(y+a_1)^3+a_0]\,r^2\,u_1} =\fft{3\,(y+a_1)^2}{2
\,[(y+a_1)^3+a_0]}. \eea
Obviously, we can set $a_1=0$ by coordinate transformation. Then the
metric is given by
\bea ds^2 &=& -\frac{y}{2 }\,\partial_r \rho\,
(\fft{dr^2}{r\,W^{\fft12}} + \fft{1}4 \,r\,W^{\fft12}\, \sigma_3^2)
-\frac{y}{4 }\rho\, (W^{\fft12}\,\sigma_1^2 +
W^{-\fft12}\,\sigma_2^2) \cr&& + \fft{3\,y^2
}{2 \,(y^3+a_0)}\, dy^2 + \fft{2 \,(y^3+a_0)}{3\,y^2
}\, (d\alpha
 -\frac{1}{4 }\rho \,\sigma_3)^2\,,\eea
where
\bea \rho&=&r\,W^{\fft12}\,\partial_r\log  u_1. \eea
Supposing
\bea
r\,W^{\fft12}\partial_r=\xi\partial_\xi=V(\rho)\,\partial_\rho\,,
\eea
we find
\bea
\partial_\rho\log\xi=\fft1V\,,~~~~ r^2=\fft{4\xi^4+a^4}{4\xi^2}\,,
~~~~W^{\fft12}=\fft{4\xi^4-a^4}{4\xi^4+a^4}\,.
\eea
Taking $\rho$ instead of $r$ as the radial coordinate, the metric
becomes
\bea\label{ehsol1} ds^2 &=& -\frac{y}{2 }\, (\fft{d\rho^2}{V}
+ \fft{V}4 \, \sigma_3^2)  -\frac{y}{4 }\rho\,
(W^{\fft12}\,\sigma_1^2 + W^{-\fft12}\,\sigma_2^2) \cr&& +
\fft{3\,y^2 }{2 \,(y^3+a_0)}\, dy^2 +
\fft{2 \,(y^3+a_0)}{3\,y^2 }\, (d\alpha
 -\frac{1}{4 }\rho\, \sigma_3)^2\,.\eea
The first equation in (\ref{ehrpoly}) becomes
\bea\label{ehrbasic} &&3
\,u_1=\fft{V}{2\,r^4\,W^{\fft12}}\partial_\rho(\rho^2)
=\fft{16\,\xi^4\,\rho\,V}{16\,\xi^8-a^8}
\,.\eea
Thus
\bea
\rho=V\,\partial_{\rho}\log u_1
=\fft1{\rho}\partial_\rho{
\left(\rho\,V\right)}-4\,
\fft{16\,\xi^8+a^8}{16\,\xi^8-a^8}\,. \eea
Then we have
\bea &&\fft {16\,\xi^8}{a^8} =\fft{\fft
1{\rho}\partial_\rho{\left(\rho\,V\right)}-\rho+4}{\fft
1{\rho}\partial_\rho{\left(\rho\,V\right)}-\rho-4} \cr&\Rightarrow&
\fft8V =\partial_{\rho}\log\left(\fft{\fft
1{\rho}\partial_\rho{\left(\rho\,V\right)}-\rho+4}{\fft
1{\rho}\partial_\rho{\left(\rho\,V\right)}-\rho-4}\right)
=-\fft{8\,\partial_\rho\left(\fft
1{\rho}\partial_\rho{\left(\rho\,V\right)}\right)-8}{\left(\fft
1{\rho}\partial_\rho{\left(\rho\,V\right)}-\rho\right)^2-16}\,. \eea
Let 
\bea \tilde V=\rho V-\fft13\rho^3\,,~~~\tilde\rho=\rho^2\,, \eea
the above equation can be rewritten as 
\bea (\tilde
V+\fft13{\tilde\rho}^{\fft32})\,\partial_{\tilde\rho}^2\tilde
V+(\partial_{\tilde\rho}\tilde V)^2-4=0\,. \eea
Supposing 
\bea \tilde V=b_3 \tilde\rho^{\fft32}+ b_2 \tilde\rho+  b_1
\tilde\rho ^{\fft12}+b_0\,, \eea 
the corresponding solutions are given by 
\bea 1)~~~\tilde V=\pm2 \,\tilde\rho+b_0\,,~~~~~~ 2)~~~\tilde V=-48
\,\tilde\rho^{\fft12}\,,~~~~~~ 3)~~~\tilde V=-\fft1{12}
\,\tilde\rho^{\fft32}-16\,\tilde\rho^{\fft12}\,. \eea 
The first solution corresponds to the flat base case discussed in the
previous section. The second solution gives rise to the metric
\bea ds^2 &=& \frac{y}{2 }\,
\left(\fft{3\,d\rho^2}{12^2-{\rho}^{2}} + \fft{12^2-{\rho}^{2}}{12}
\, \sigma_3^2\right) \cr&&+\frac{y}{4 }\,
\left[\left({12-\sqrt{12^2-\rho^2}}\right)\,\sigma_1^2
+\left({12+\sqrt{12^2-\rho^2}}\right)\,\sigma_2^2\right] \cr&& +
\fft{3\,y^2 }{2 \,(y^3+a_0)}\, dy^2 +
\fft{2 \,(y^3+a_0)}{3\,y^2 }\,\left (d\alpha
 -\frac{1}{4 }\rho\, \sigma_3\right)^2\,.\eea
With certain coordinate transformation and rescaling of the metric,
it can be expressed as 
\be ds^2= \fft{dy^2}{W} + \ft14 W y^2 (d\alpha - 2 \sin(2\rho)\,
\sigma_3 )^2 + y^2 ds_{\CP^2}^2\,. \ee
where $W$ is given by (\ref{wv}) and $ds_{\CP^2}^2$ is the triaxial
$\CP^2$ metric given by (\ref{cp2s2s2}). Thus, the metric describes
an $\R^2\times \CP^2$ instanton that is asymptotic to the
$\R^6/Z_3$.

  The third solution gives the metric
\bea ds^2 &=& \frac{y}{2 }\,
\left(\fft{4\,d\rho^2}{8^2-{\rho}^{2}} + \fft{8^2-{\rho}^{2}}{16} \,
\sigma_3^2\right) +\frac{y}{4 }
\left(\fft{\rho^2}{8}\,\sigma_1^2 +8\,\sigma_2^2\right) \cr&& +
\fft{3\,y^2 }{2 \,(y^3+a_0)}\, dy^2 +
\fft{2 \,(y^3+a_0)}{3\,y^2 }\, \left(d\alpha
 -\frac{1}{4 }\rho\, \sigma_3\right)^2\,.\eea
With certain coordinate transformation and rescaling of the metric,
it can be expressed as 
\be ds^2= \fft{dy^2}{W} + \ft14 W y^2 (d\alpha -\ft43 \sin\rho\,
\sigma_3)^2 + \ft13 y^2 ds_{S^2\times S^2}^2\,. \ee
where $W$ is given by (\ref{wv}) and $ds_{S^2\times S^2}^2$ is the
triaxial $S^2\times S^2$ metric given by (\ref{cp2s2s2}). Thus, the
metric describes an $\R^2\times S^2\times S^2$ instanton that is
asymptotic to the cone over $T^{1,1}/Z_2$.

\section*{Acknowledgement}

     We are grateful to Chris Pope for useful discussion.
Z.L.W. acknowledges the support by grants from the Chinese Academy of
Sciences, a grant from 973 Program with grant No: 2007CB815401 and
grants from the NSF of China with Grant No:10588503 and 10535060.

\end{document}